\documentclass[10pt]{article}
\usepackage[T1]{fontenc}
\usepackage{graphicx}
\usepackage{amsmath}
\usepackage{hyperref}
\usepackage{authblk}

\begin{document}

\title{Holographically controlled three-dimensional atomic population patterns}
\author{Adam Selyem,$^{1,*}$ Sylvain Fayard,$^2$ Thomas W. Clark,$^{1,3}$ Aidan S. Arnold,$^{4}$ Neal Radwell,$^{1}$ and Sonja Franke-Arnold$^{1}$}
 \affil{$^{1}$School of Physics and Astronomy, SUPA, University of Glasgow, G12 8QQ, Glasgow, United Kingdom\\
$^{2}$Institut d'Optique Graduate School, 91127 Palaiseau, France\\
$^{3}$Institute for Solid State Physics and Optics, Wigner Research Centre for Physics, Budapest, Hungary\\
$^{4}$Department of Physics, SUPA, University of Strathclyde, G4 0NG, Glasgow, United Kingdom}

\maketitle

\begin{abstract}
The interaction of spatially structured light fields with atomic media can generate spatial structures inscribed in the atomic populations and coherences, allowing for example the storage of optical images in atomic vapours.  Typically, this involves coherent optical processes based on Raman or EIT transitions.  Here we study the simpler situation of shaping atomic populations via spatially dependent optical depletion.  Using a near resonant laser beam with a holographically controlled 3D intensity profile, we imprint 3D population structures into a thermal rubidium vapour.  This 3D population structure is simultaneously read out by recording the spatially resolved fluorescence of an unshaped probe laser.  We find that the reconstructed atomic population structure is largely complementary to the intensity structure of the control beam, however appears blurred due to global repopulation processes.  We identify and model these mechanisms which limit the achievable resolution of the 3D atomic population.  We expect this work to set design criteria for future 2D and 3D atomic memories. 
\end{abstract}


\section{Introduction}

The drive towards communication systems with increasing data capacity and density has led to an interest in utilising the spatial degree of freedom as additional information career.  For optical communications this is most readily achieved with spatial light modulators (SLMs) and digital micromirror devices (DMDs), displaying computer-generated holograms that control the phase, intensity and even polarisation profile of laser beams \cite{Roadmap,OPN2017,Igasaki1999,Chen2011,Zhu2014,Rong2014,Clark2016,Bowman2017,Mitchell2017}.  Shaped laser beams are of interest in a number of different contexts including quantum communication \cite{Trichili2016, Willner2015}, optical trapping of microparticles \cite{Grier2003,Lin2012} as well as cold and ultracold atoms \cite{Ketterle1993,Dholakia1998,Schiffer1998,Radwell2013b,Lembessis2015}.

While advanced techniques already exist for the generation of spatial light structures, the development of appropriate atomic interfaces is paramount for image based quantum communication networks.  
Recent experiments have demonstrated the control \cite{Walker2012,Bouchard2016} and storage of two-dimensional quantum images in room temperature \cite{Shuker2008} and cold \cite{Ding2013,Veissier2013,Nicolas2013} atomic gases.  A related technique based on wavevector multiplexing presents an alternative promising atomic memory \cite{Parniak2017}, and the shaping of quantum degenerate gases has been demonstrated as an enabling tool for atomtronic applications \cite{Gauthier2016}.  

In this work we demonstrate the mapping of 3D intensity structures onto 3D atomic population structures via spatially dependent optical depletion, and read out the 3D population distribution in a technique similar to electron shelving \cite{Nagourney1986,Sauter1986,Bergquist1986}.   This extends the concept of 2D images to 3D sculptures, for both light and atomic medium.  Access to the 3rd dimension effectively reveals 2D phase information, so that our (incoherent) 3D intensity mapping may offer an alternative to 'traditional` quantum memories, storing 2D phase and amplitude information.  As amplitude information tends to be more robust than phase information \cite{Hau2001} this may increase possible storage times.

\section{Method}

By defining the phase and amplitude of a light field in the plane of an SLM, its  propagation in 3D is completely determined. For most applications it is the light profile in a single plane that is of interest, however here we are concerned with the full 3D structure of the light. A wide range of 3D structures can be realised, as long as they obey Maxwell's equations \cite{Shabtay2003}, and more complicated light sculptures could be generated by using the interference between multiple (e.g. counterpropagating) beams \cite{Arlt2000,Leach2005,Courtial2006,Arnold2012,Cameron2017}.

\begin{figure}
	\centering
	\includegraphics[width=.9\linewidth]{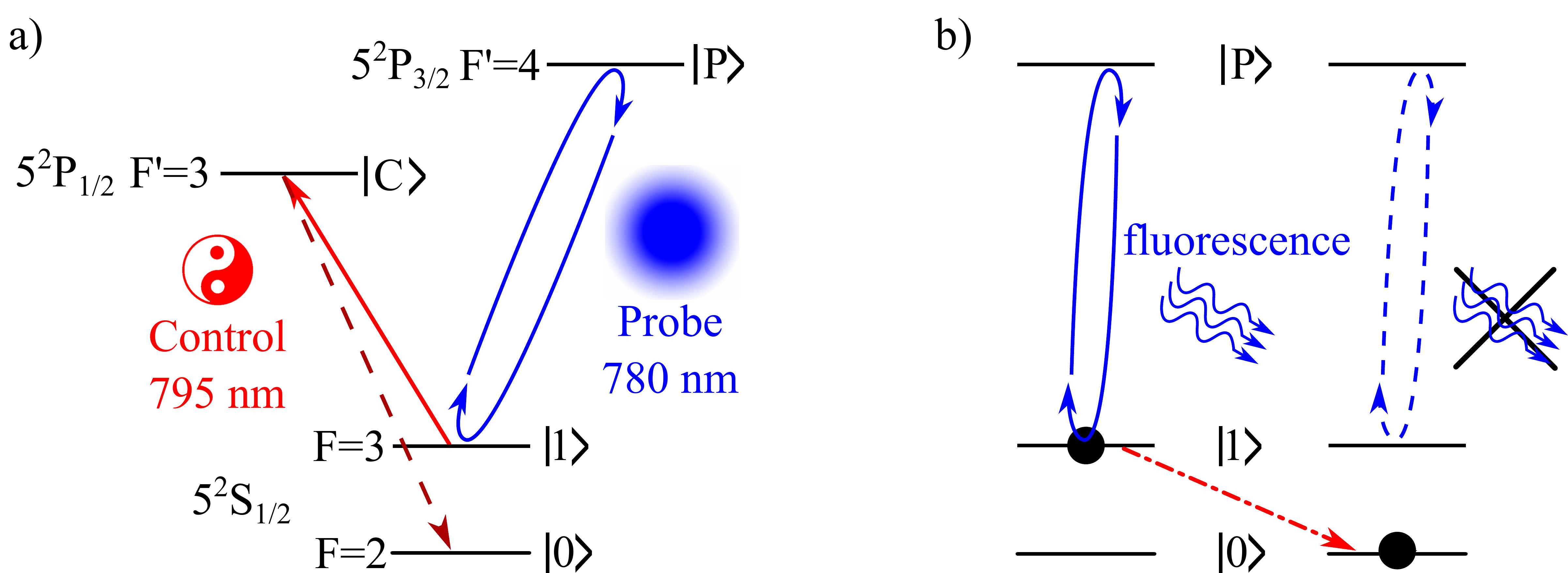}
	\caption{\label{level_scheme_fig}a) Simplified energy level scheme of Rb$^{85}$ showing spectroscopic notation. D$_1$ and D$_2$ transitions are shown in red and blue, respectively. b) Interpretation in terms of electron shelving. Detecting fluorescence of the probe light reveals that the atom is in $|1\rangle$, the absence of fluorescence indicates that the atom was transferred from $|1\rangle$ to $|0\rangle$ due to excitation by the control beam and subsequent spontaneous decay.}
\end{figure}

We use a 3D structured light beam to shape the local population distribution of an atomic vapour via optical depletion.  We can read out this generated population structure from the fluorescence of an unstructured probe beam, as shown in Fig. \ref{level_scheme_fig}a).   More specifically, we encode spatial information as population imbalance between the two hyperfine ground states $5^2S_{1/2}$ F=2 (denoted as $|0\rangle$), and  $5^2S_{1/2}$ F=3 ($|1\rangle$) in rubidium-85.  The continuous-wave control laser is tuned to the D$_1$ transition at 795~nm, and shaped holographically in phase and amplitude by an SLM. Population is transferred from $|1\rangle$ via $|C\rangle$ to $|0\rangle$ by resonant excitation with the control laser and subsequent spontaneous decay with a branching ratio of $\approx 2:1$.  We detect the remaining population in $|1\rangle$ by observing the fluorescence of the continuous-wave probe laser tuned to the D$_2$ transition at 780~nm.   Note that the probe beam drives a cycling transition, and hence does not transfer atoms between ground states except through spurious off-resonant pumping, which occurs at a rate three orders of magnitude slower than the probe transition.  
The observed fluorescence (or absence thereof) indicates that the atom is in state $|1\rangle$ (or state $|0\rangle$), as indicated in Fig. \ref{level_scheme_fig}b) . 

This procedure can be understood as a 3D analogue to electron shelving, a ubiquitous technique in ion and atom trapping which can probe the quantum state of an individual atom by testing the fluorescence on a cycling transition. More specifically, an atomic quantum state inscribed as a superposition of two ground or meta-stable states $|0\rangle$ and $|1\rangle$ may be determined by inducing Rabi oscillations between $|1\rangle$ and a short-lived excited state $|P\rangle$ with a probe laser, and monitoring its fluorescence.  Unlike typical electron shelving applications we do not, of course, generate coherent superpositions as our population imbalances are due to optical pumping.

\par
Initially, in the absence of light, the atoms are distributed between the hyperfine ground states $|0\rangle$ and $|1\rangle$ at a ratio of 5:7 according to their degeneracy, $P_0 = 5/12$ and $P_1 = 7/12$.
In the presence of the control beam, the local population distribution is altered as a result from two competing processes.  The number of atoms in $|1\rangle$ is depleted depending on the local intensity of the control laser, and it is repopulated by atoms drifting back into the observation region after collisions with the cell walls. We now consider these two processes in more detail: 

Atoms in high intensity regions of the control beam are transferred rapidly from $|1\rangle$ to $|0\rangle$ where they no longer interact with either the control or probe light. Conversely, in regions where the control beam has zero intensity, atoms remain in $|1\rangle$. The populations could be derived from spatially dependent optical Bloch equations, but for our purposes it suffices to consider optical pumping in terms of rate equations. The spatially dependent depletion rate of the upper ground state $|1\rangle$ is given by $ -\frac{d}{dt} P_1 =\frac{d}{dt} P_0 = R_{\rm c} P_1$, where $R_{\rm c}$ is the scattering rate of the control laser. In a warm atomic gas, $R_{\rm c}$ has been derived for a 2-level atom \cite{Corney1977}, and more recently was experimentally observed in the presence of optical pumping \cite{Siddons2008}, but in both cases without considering any spatial dependence. In our case the scattering rate is 

\begin{equation}
\label{scatteringEquation}
R_{\rm c}(\mathbf{r}) = \frac{2}{6}\frac{\Gamma \lambda}{4 \sqrt{3 \pi k_{\rm b} T / m_{\rm Rb85}}}\frac{\Gamma}{2}\frac{I(\mathbf{r})/I_{\rm S}}{\sqrt{1+I(\mathbf{r})/I_{\rm S}}},
\end{equation}
where $I(\mathbf{r})$ is the local intensity of the control beam, $T \simeq 293$~K the temperature in the cell, $m_{\rm Rb85}$ is the mass of a rubidium 85 atom, $\lambda = 795$~nm the wavelength, $I_{\rm S} = 4.49$~mW/cm$^2$ the saturation intensity and $\Gamma =  2 \pi \times 5.75 $~MHz \cite{Steck} the decay rate. The leading factor of 2/6 comes from the branching ratio and relative pumping rate between the probe and control beams, reflecting the probability for an atom to decay to $|0\rangle$ rather then back into $|1\rangle$.
\par
For cold, stationary atoms even a minute intensity of the control laser would lead over time to a complete depletion of $P_1$.   For the moving atoms in our vapour, the 
local population $P_1$  is constantly replenished due to atoms drifting into the observation region from outside of the control beam.  Since the density of the gas within the cell is low (on the order of 10$^{-8}$ Torr), the mean free path between atom-atom collisions is on the order of kilometres and atoms travel along straight lines. 
Collisions with the cell wall 'reset` the atomic populations.   On the path from the cell wall towards the local observation region atoms may need to traverse the control light field and therefore may be partially depleted before reaching the observation region.  While depletion is dictated by the local control laser, repopulation is a more complex process that depends on the global control beam structure as well as the transverse velocity of the atoms.  
\par
In order to calculate the overal repopulation rate it is sufficient, to first order, to consider atoms travelling in a plane transverse to the beam propagation.  As our control laser is resonant with stationary atoms, any axial velocity will lead to a Doppler shift that reduces the optical pumping rate. Detecting the population $P_1$ requires both control and probe laser to address the same velocity class of atoms within a narrow detuning band which we measured to be approximately 40 MHz wide, corresponding roughly to the convolution of the probe and pump line shapes. Considering atoms travelling within the plane perpendicular to the light propagation along a straight path from the cell wall to the local observation region, we can express the variation of $P_1$ as a combination of depletion and repumping, as 
\begin{equation}
\frac{\textrm{d}P_1}{\textrm{d}r} = \frac{-P_1 R_{\rm c}}{v_r},
\end{equation}
where $v_r = \sqrt{2k_bT/m_{Rb85}}$ is the most probable transverse speed of atoms. Integrating this equation, considering all the directions an atom can travel along and extending to 2D one finds that the $|1\rangle$ population at a position $(x,y)$ in a plane transverse to the beam propagation can be expressed as
\begin{equation}
\label{depopulationEquation}
\begin{split}
P_1(x,y) & \propto \frac{7}{12} \int\limits_{0}^{2\pi}\textrm{d}\theta P_{1,\theta}(\theta),  \\
\textrm{with} \, P_{1,\theta}(\theta) &= \exp \left( \int\limits_0^{wall} \textrm{d}r \frac{- R_{\rm c}(r \cos \theta + x, r \sin \theta + y)}{v_r}\right).
\end{split}
\end{equation}
Here $\theta$ and $r$ are polar coordinates defined with respect to the origin at $(x,y)$ and the prefactor of $7/12$ is the initial $P_1$ population. The result of these processes is that the population in the upper ground state acquires a spatial structure that is inverse to the control beam shape, modified due to globally sampled repopulation.
We note that the derivation of \ref{scatteringEquation} takes into account the full Maxwell-Boltzmann distribution, but we find using the most likely transverse speed $v_r$  in \ref{depopulationEquation} suffices to simulate our experimental observations.
\par

We verify the 3D intensity structure of our control beam as well as the resulting 3D population structure from spatially resolved fluorescence measurements of the shaped control beam and an unshaped probe beam resepctively.  We have recently shown that a 3D light structure can be tomographically reconstructed by observing its fluorescence from a (uniform) atomic vapour \cite{Radwell2013}. 
In our previous work, and here for the verification of the control beam structure, the atoms simply provide a passive scattering medium.  
For our current work we are instead interested in the 3D population structures which arise as a result of optical pumping between the atomic levels. We can reconstruct this 3D atomic population distribution pattern by observing the fluorescence of a uniform probe beam at a different frequency.  

The probe beam drives the cycling transition between $|1\rangle$ and $|P\rangle$ as shown in  Fig. \ref{level_scheme_fig}b), and co-propagates through the rubidium vapour cell with the control beam. Its profile is a truncated Gaussian, ensuring a relatively uniform beam intensity across the observation region.  The probe beam fluorescence is proportional to $P_1 R_{\rm p}$, where $P_1$ is the spatially varying population of  $|1\rangle$ and $ R_{\rm p}$ is the scattering rate of the probe beam.  This scattering rate is identical in form to $R_{\rm c}$ in Eq.~\ref{scatteringEquation}, a saturation intensity $I_{\rm S} = 3.90$~mW/cm$^2$, decay rate $\Gamma =  \pi \times 6.07\,$ ~MHz and a prefactor of 1 corresponding to the branching ratio for this cycling transition. By recording the spatially varying fluorescence and applying the tomographic techniques outlined in \cite{Radwell2013} we can therefore non-destructively map out the 3D population structure of $P_1$.  In our case, for a relatively uniform probe beam intensity, the probe beam fluorescence  corresponds directly to the population in $|1\rangle$, but more generally could also be inferred from a reconstruction using based on a non-uniform probe beam with a known profile.

\section{Experimental results}

\begin{figure}[!b]
	\centering
	\includegraphics[width=.45\linewidth]{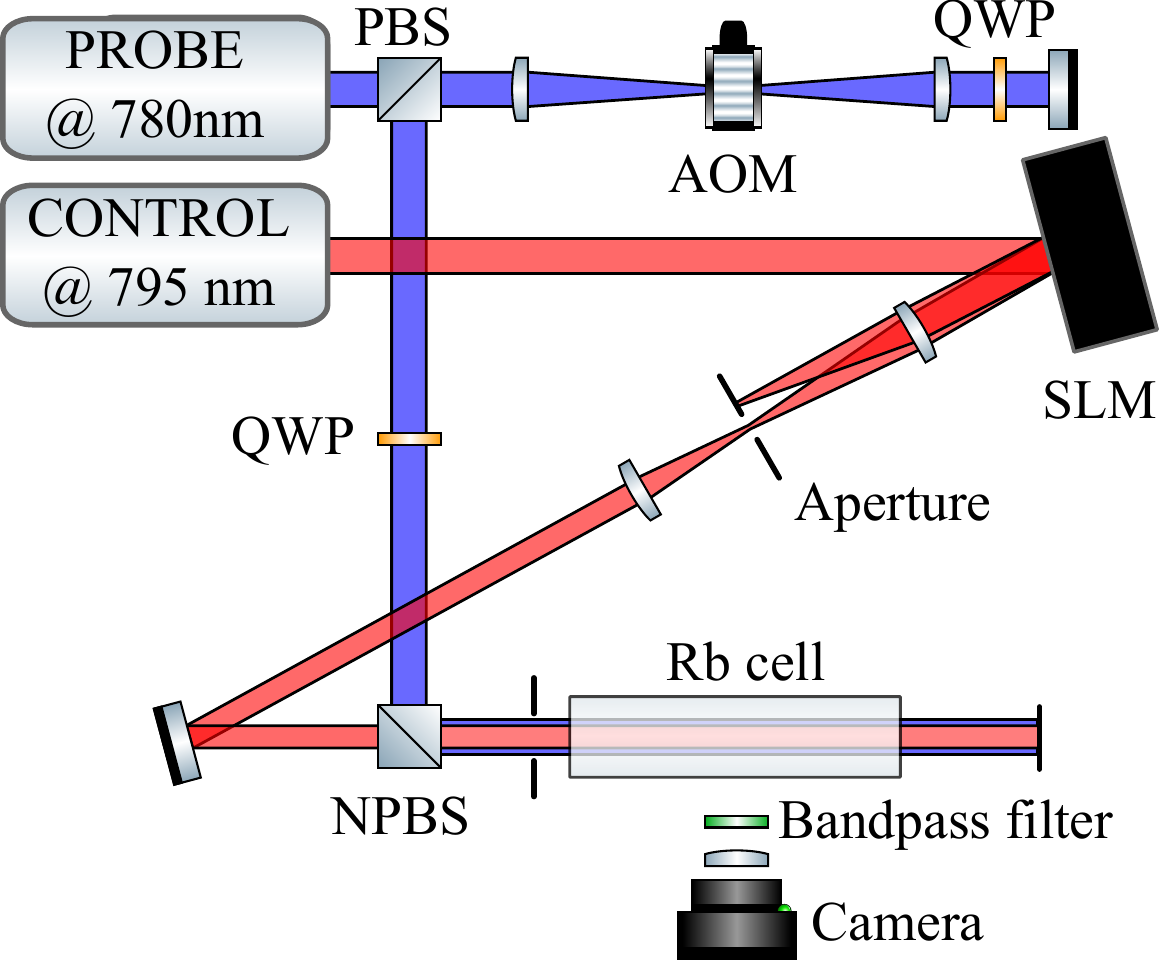}
	\caption{\label{expt_setup_fig}Schematic diagram of the experimental setup. AOM: acousto-optic modulator, QWP: quarter waveplate, (N)PBS: (non-) polarising beam splitter, SLM: spatial light modulator.}
\end{figure}

The experimental setup is shown in Fig. \ref{expt_setup_fig}. Both home-made external cavity diode lasers are locked to their respective transitions via Doppler-free spectroscopy. For fine frequency control the probe beam passes through an acousto-optic modulator (AOM) in double-pass configuration. The control beam is diffracted off an SLM (Hamamatsu LCOS X10468) displaying a computer-generated hologram \cite{Clark2016}. The diffraction order containing the shaped beam is selected by an aperture in the Fourier plane of the hologram. The control and probe beams are expanded to a diameter of approximately 10 mm and 12 mm respectively and combined on a non-polarising beam splitter to co-propagate through a vapour cell containing thermal rubidium. Both beams pass through a circular aperture, the size of which is chosen such that the control beam barely passes through and the probe beam is truncated to approximate a flat top beam. This is done to limit probe beam fluorescence from regions containing no information, since this fluorescence would reduce the contrast of the recorded images. The vapour cell is imaged from the side by a camera through a bandpass filter centred on 780 nm with a bandwidth of approximately 10 nm. The imaging lens (f=16mm, NA=0.6) was selected such that the depth of field is sufficient to sharply image the whole beam volume.

\begin{figure}
	\centering
	\includegraphics[width=.9\linewidth]{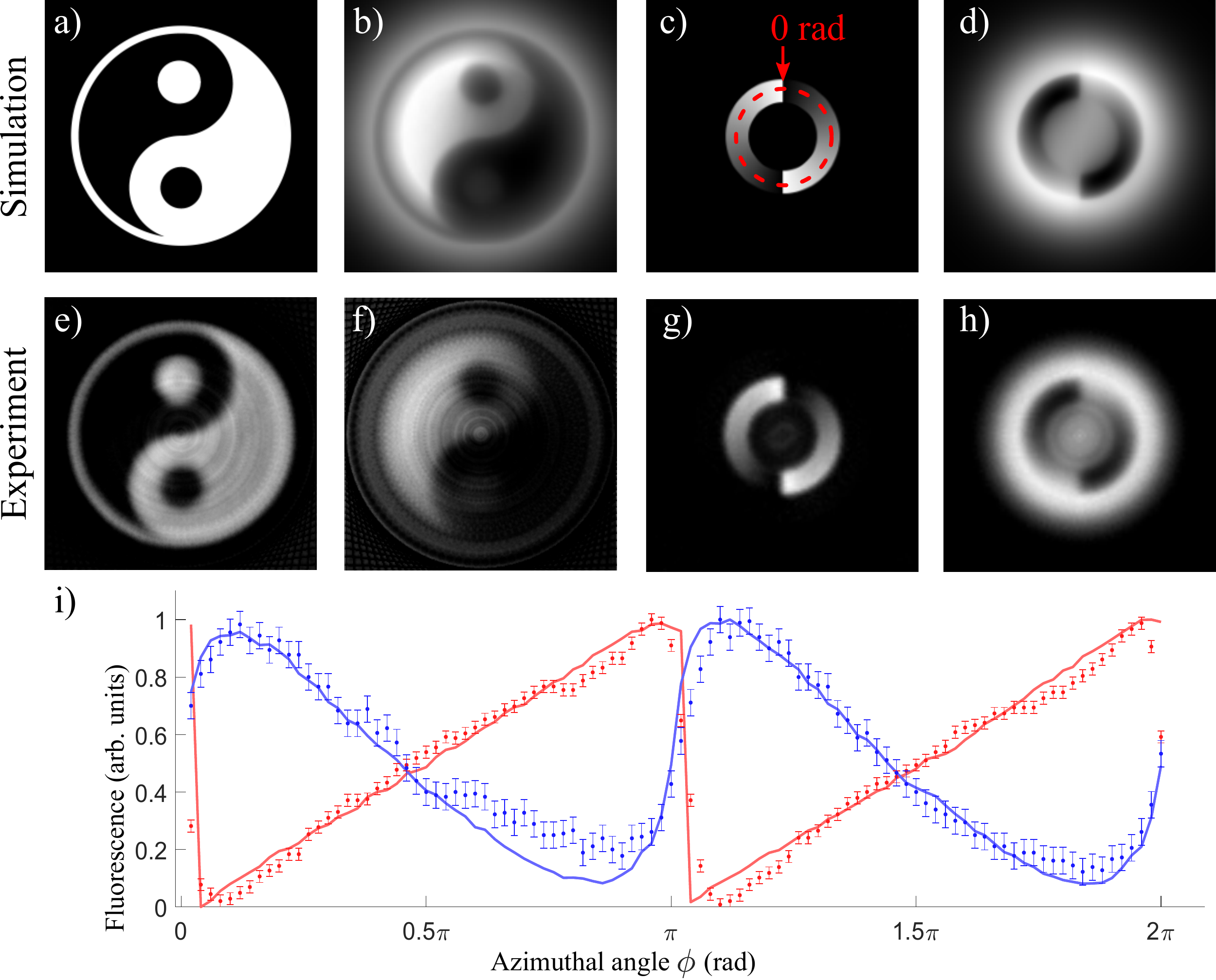}
	\caption{\label{yinyangresults_fig}Simulated and reconstructed cross-sections from a control beam with a yin-yang symbol profile (a, b, e, f) and a linear azimuthal intensity ramp (c, d, g, h), measured in the image plane of the SLM. a,c) Desired control intensity profile. e,g) Corresponding reconstruction from fluorescence of the control beam. b,d) Simulation of probe fluorescence indicating population in $|1\rangle$. f,h) Measured probe fluorescence reconstruction. i) Unwrapped control (red) and probe (blue) fluorescence profiles, individually peak-normalised, at a fixed beam radius corresponding to the red dashed circle in c).  Error bars represent the standard deviation calculated from 10 cross-sections and solid lines are simulations.}
\end{figure}

We demonstrate the atomic depletion due to different control beam intensities in two experiments, one with a cross-section in the form of a yin-yang symbol, shown in Fig.~\ref{yinyangresults_fig}a), and one with a ring profile with azimuthally increasing intensity, in Fig.~\ref{yinyangresults_fig}c). In each case we report the simulated and measured fluorescence of the control beam, indicating the optical information, and the resulting fluoresence of the unshaped probe beam, indicating the resulting population structure. In order to compare the measured profile to our model we numerically evaluated Eq. \ref{depopulationEquation} using the ideal control beam profile and a truncated Gaussian probe beam profile, which matched our measured probe beam profile.
For simulated (Fig. \ref{yinyangresults_fig} a) and b), e) and f)) as well as for observed beam profiles (Fig. \ref{yinyangresults_fig} e) and f), g) and h))  we see that the control and probe fluorescence are complementary to each other, indicating a controlled population transfer from $|1\rangle$ to $|0\rangle$.  Note that in dark regions of the control beam which are surrounded by light, the remaining population in $|1\rangle$ is highly reduced, even though there is no control light in these areas. This limits the range of population patterns that can be inscribed onto a thermal vapour by limiting the contrast achievable between high and low population regions. This effect is reproduced in our simulations, with good agreement between the expected and measured cross-sections.  This can be seen more quantitatively in Fig.~\ref{yinyangresults_fig}i), showing the polar plots for the ring shaped control beam. The population distributions (shown in blue) mimick the inverted control light profile (shown in red), but appear blurred.   As mentioned before, we believe that the blurring originates from the repopulation rate which samples the global control beam structure.   The blurring depends implicitly on various factors, included in our model via equations \ref{scatteringEquation} and \ref{depopulationEquation}, which in combination limit the resolution of population patterns that can be imprinted onto a thermal vapour.  These include the intensity profile of the control beam and positioning within the vapour cell, its detuning, the temperature and velocity profile of the atoms, and of course the excitation and decay channels determined by the specific atomic level structure.  

\begin{figure}
	
	\includegraphics[width=.95\linewidth]{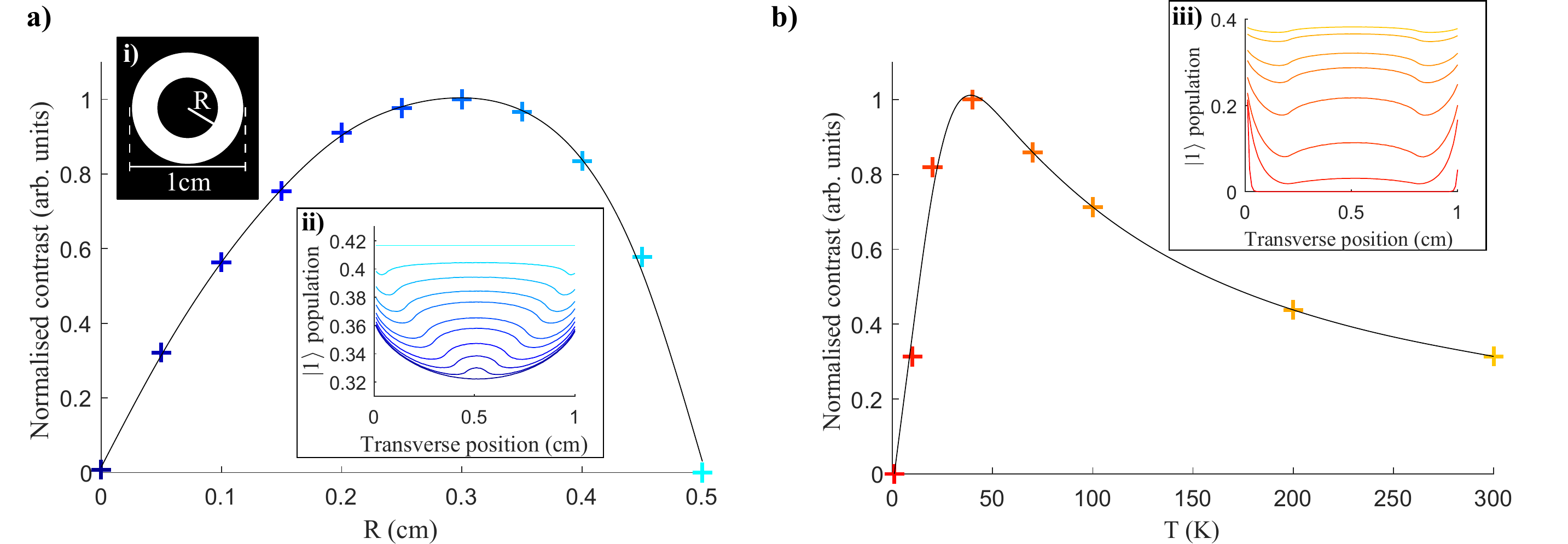}
	\caption{\label{contrastData_fig}Normalised population contrast as a function of varying a) dark core radii R in a flat-top control beam and b) temperature. In a) radii R are ranging from 0~cm (shown in dark blue) to the full beam radius of 0.5~cm (cyan) at a temperature of 300~K. In b) the temperature is ranging from 1~K (shown in red) to 300~K (yellow), for R=0.3. Fits show as black lines are smoothed spline interpolations calculated from 15 data points in both figures. Inset i) shows the control beam shape, indicating the definition of R. Insets ii) and iii) show population pattern cross-sections through the centre of the control beam.}
\end{figure}

Our model allows us to demonstrate the influence of different factors on the expected blurring, the most important being beam shape and atomic temperature, shown in Fig. \ref{contrastData_fig}a) and b) respectively. We investigated how the population contrast changes at the position of a dark core which is surrounded by a varying amount of light intensity.   We simulated the population patterns generated by a flat-top control beam with a central dark core of varying radius $R$, where the outer beam radius was fixed at 0.5~cm and the local intensity was $0.1\,I_{\rm S}$. We define the contrast as the population $P_1$ in the beam centre minus the minimum population within the bright control beam area.   Without a hole, there is of course no contrast, and we see that the population $P_1$ in the centre is reduced compared to the outer sections of the beam, as fewer undepleted atoms reach the beam centre. With increasing dark core radii, the contrast initially increases until it peaks at a radius of approximately 0.3~cm. For larger dark core radii the contrast decreases again because the amount of control light is not sufficient to effectively pump atoms into $P_0$ during the transit across the bright ring. Similar effects are at work also for more complicated beam shapes, including the yin-yang pattern and azimuthal intensity pattern discussed in Fig. \ref{yinyangresults_fig}.

Fig~\ref{contrastData_fig} b) shows simulations using the control beam of \ref{contrastData_fig} a), with a fixed core radius of R= 0.3~cm, at various atomic temperatures. We find that at very low temperatures, below approximately 20~K, the contrast is low. This is because the slow-moving cold atoms will spend a long time within the control beam and will be very efficiently transferred to $|0\rangle$, so  that atoms in $|1\rangle$ do not reach the dark core. For higher temperatures, {\it i.e.}~above around 100~K, contrast is low again because the fast atoms do not spend sufficient time in the control beam to be transferred to the lower ground state. There appears to be an optimal temperature range, close to 50~K, where maximum contrast can be achieved for the beam parameters used here. We note, however, that for a different set of parameters, such as intensity and beam shape, this optimal temperature will take a different value.  We also note that all our experiments and simulations are based on CW control and probe beams, and don't reflect depletion processes in applications using a pulsed control beam. 

\begin{figure*}[!t]
	\centering
	\includegraphics[width=1 \columnwidth]{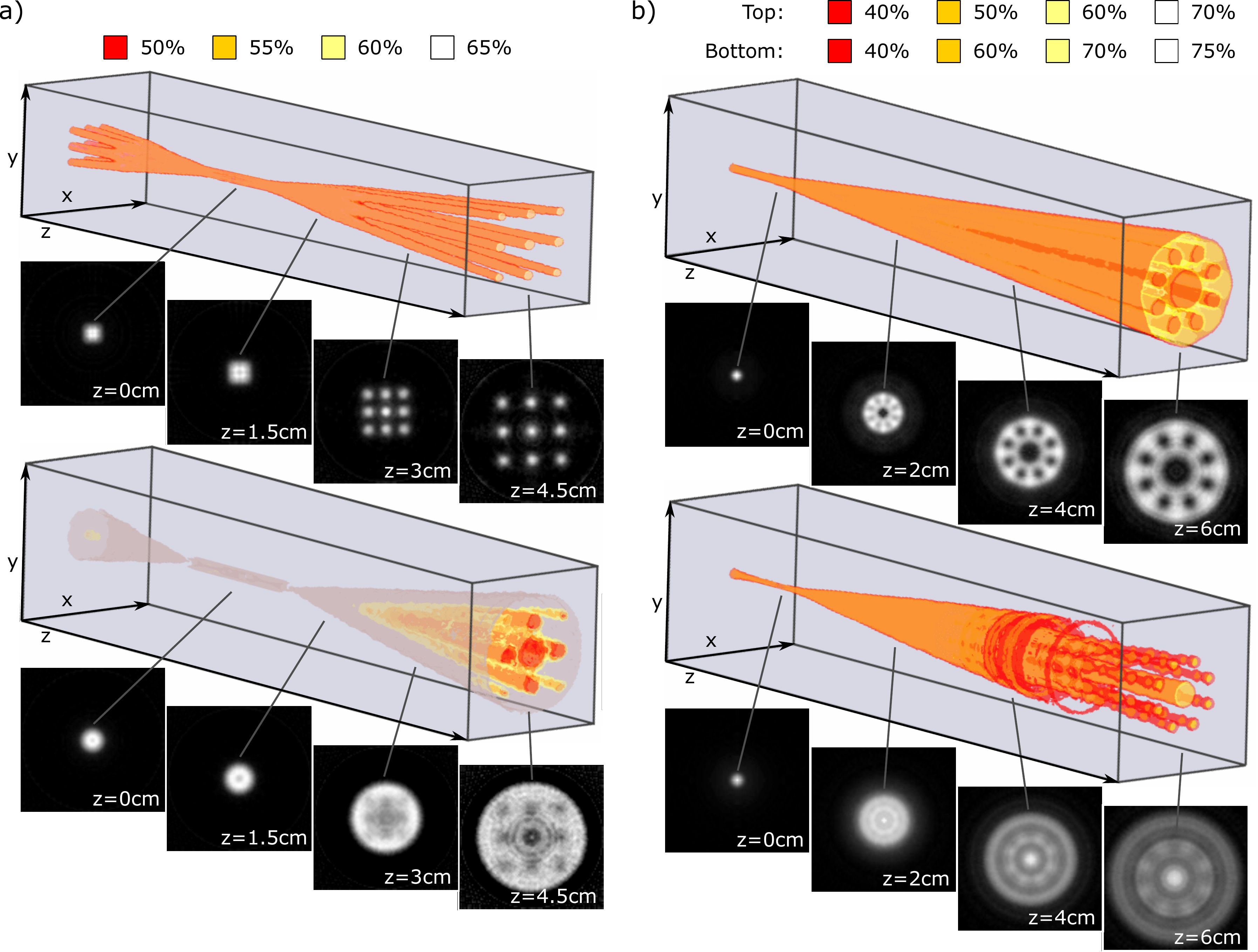}
	\caption{\label{3Dresults1_fig}Full 3D reconstructions of the atomic population structure arising from depletion via structured control light beams.  The control beams had a cross-section of a) a 3-by-3 array of discs and b) an optical Ferris wheel,  propagating in the $-z$ direction. Observing the scattering of the control beams  allows us to visiualise the light field structure (top panels).  Observing the scattering of an unshaped probe beam allows us to reconstruct the resulting population structures in $|1\rangle$ (bottom panels).  
The dimensions of the boxes are approximately 1 cm x 1 cm x 7 cm. Propagation distances $z$ are measured from the focus. For clarity, isosurfaces were plotted at intensities indicated in the colourboxes. For clarity we also show sample cross-sections.}
\end{figure*}

Finally we note that our method allows us full access to the 3D structures of the control light and the resulting imprinted population structures. In Fig.~\ref{3Dresults1_fig} we show full 3D structures of control and probe fluorescence. The control beam cross-section used in Fig. \ref{3Dresults1_fig}a) was a 3-by-3 grid of discs. Both the control and probe beams were focussed by a lens with focal length 50 mm placed immediately before the vapour cell. In Fig. \ref{3Dresults1_fig}b) the control beam was an optical Ferris wheel \cite{Franke-Arnold2007}, focussed by a 75 mm lens. The 3D structures are plotted as isosurfaces constructed from 175 peak-normalised cross-sections. As can be seen in the sample probe fluorescence cross-sections the effects discussed above have varying degrees of importance on different scales. For large beams and relatively low intensities the population pattern approximates the inverse of the control profile reasonably well. However, as intensities increase and the beam sizes decrease the blurring effect leads to a loss of contrast, and the imprinted patterns differ from the inverted control beam.

It is worth noting that scattering rates are sensitive to light polarisation which we do not take into account here. However, since we do not control magnetic fields at the detection cell the magnetic field of the Earth sets the quantisation axis for the atoms, relative to which polarisation axes are defined in the frame of the atoms. We find that this magnetic field has components along the $x, y$ and $z$ axes defined by the polarisation of our lasers and their propagation direction,and as a consequence the polarisation of the lasers in the experiment have no strong effect.

\section{Conclusion}
We have demonstrated the inscription of three-dimensional atomic population structures in two hyperfine ground states of rubidium. We achieve this by depleting the upper hyperfine ground state with intensity structured light, and we read out the resulting population structure by spatially resolved fluorescence measurements.  We find that the achievable resolution of the population pattern depends on the interplay of local depletion through the control beam and repopulation that is influenced by the global shape of the control beam.  We model these processes using rate equations and find good agreement between simulated and observed population patterns.  Our work indicates that optical pumping and repopulation play an important role for the propagation of spatially structured light through atomic media, and similarly for the storage of images or 3D light patterns within atomic gases.  

In our work, the light structures as well as the atomic population structures represent 3D digital (i.e. not complex) information.  It is interesting to note that access to 3D intensity information effectively gives access to 2D phase information, and our suggested technique may offer an alternative path to access complex 2D information.  
More generally, however, our method could also be extended to phase-coherent imprinting techniques for true quantum systems, storing 3D \emph{quantum} information in quantum degenerate gases or even the spatial structure of individual atoms or ions.  Our current work demonstrates a first step towards such 3D image memories, imprinting optical information into 3D population structures of a room temperature rubidium vapour, with the individual atoms act as transient qubits distributed at random positions in 3D.

\section*{Funding}
We gratefully acknowledge funding through the Leverhulme Trust (RPG-2013-386), UK Quantum Technology Hub in Quantum Enhanced Imaging (EP/M01326X/1) and the European Training Network ColOpt (H2020-MSCA-ITN-2016).

\section*{Acknowledgements}
We thank Rachel Offer, Benjamin Carr and Francesco Castellucci for helpful discussions.

\bibliography{bibliography}  
\end{document}